## ENLARGING HOLOGRAMS UNDER WHITE LIGHT, A WAY TO SAVE HOLOGRAPHIC MATERIAL

### J.J. Lunazzi

When photography became popular the most common substrate material employed were glass plates. Materials were usually of high resolution, so that it was soon after when nice projections were made in darkrooms, creating the slide projection utility, which is still employed. The saving in material was enormous, and enabled fast changing from one picture to the following. When the substrate was plastic, cinematography occured. A parallel technical comparison with the development of holographic cinematography could be useful, as follows.

Although the enlargement of holograms was the subject of many researchs (1), enlarged holographic cinematography was performed only during the period 1980-1984 in Russia (2). To our knowledge, the project was no longer developed. The cost of dealing with laser light at the projection step could be one reason for that. The holographic screen employed worked in reflection, 1 x 0,8 m wide, and two observers could see the scene at a distance of several meters.

The screen was probably rigid, designed mainly for one observer, and the method for extending the audience was the generation of different zones were the scene appears repeated.

We developed techniques for employing holographic screens under white light, first exhibited to the public in 1989 (3) (4) which demonstrated the possibility of enlarging holograms already in 1990 (5). Fig. 1 shows how we encode views in a first step. The plane of the figure is horizontal, and a thin vertical slit is necessary to project each different view of the scene at a different sequential wavelength value. Decoding happens because the projected views are naturally angularly redistributed by the holographic screen (6).

The screens we had at that time were rather small (15 cm x 30 cm) but made of embossed material working in transmission. With Xenon lamps we could demonstrate the possibility of further enlarging (7) soon reaching the format of 0.75 cm x 1.14 m on plastic substrate. Although it was of an equivalent size to the ones made at Rusia, it allowed at least six simultaneous observers seated at three different distances. No view was repeated, but only horizontal parallax was shown. The closest distance being less than one meter, a larger angular field of view resulted. Already in 1993 we showed large images on that screen, but the very practical results started in 1994 after inventing a technique for concentrating the hologram's light onto the projecting lens (8) (9). Ordinary halogeneous projection lamps can now be employed, and registering is made in 35 mm format holographic film. Because only ordinary photographic optics is employed, the size of the subjects is limited to small objects.

Fig.2 is a stereo photographic pair of the scene registered with a camcorder video camera. It is shown in the format left view - right view - left view (L-R-L) so that the observer who wants to see it stereoscopically can choose the LR sequence for parallel observation, or RL for crossed observation. A 20% of lateral compression was performed to better fit the figure into text.

Everything is ready for performing an animation, and a new step towards holographic cinematography will be given soon. Even if cinematographic examples are not ready yet, the saving in photosensitive material is so large (600 times), and because holographic film is no longer manufactured in large sizes nowadays, it becomes evident that this technique renders an important economy for large holographic displays. Because it is based in the horizontal dispersion of light, in order to allow a larger audience it will be necessary to obtain an important increase on the number of lines/mm with which the screens are made. It would also be necessary to add some new process to obtain bright diffraction at one third of the visible's spectrum bandwidth, a challenge to bring color images for the public. We hope that applications will start to be used extensively and that the technique attract the interest of the holographic community.

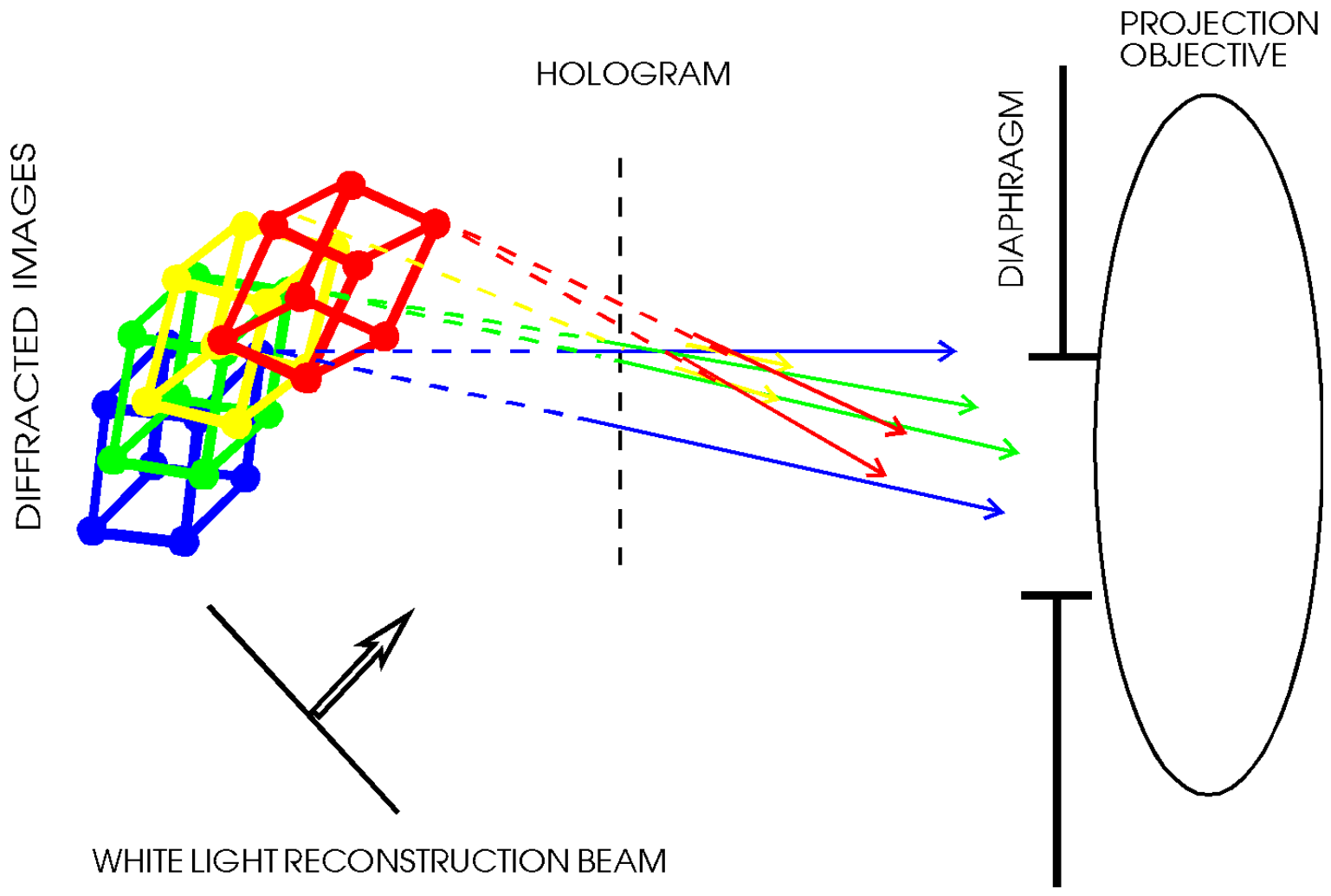

**Fig.1:** Wavelength encoding of views by a hologram, the basic principle of the technique.

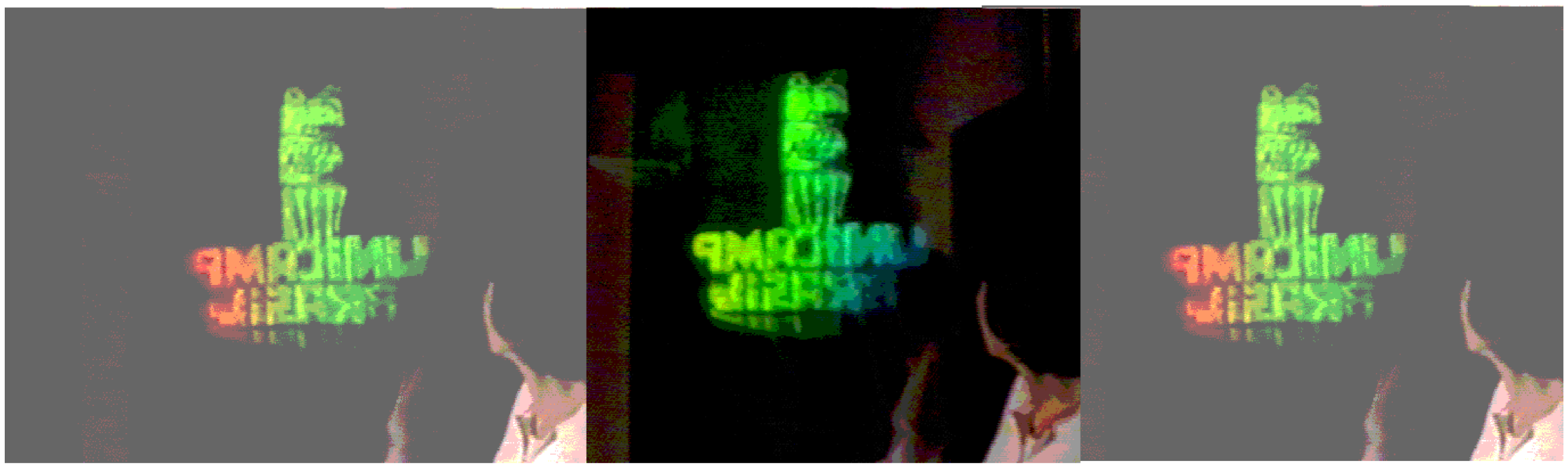

**Fig.2:** Stereoscopic pair (l-r-l) of an enlarged hologram seen by one person.

---

References:

1. E.N. Leith et al., "*Holographic cinematography",* App. Opt. 11, 9 , p. 2016-2023 (1972).
2. V.G. Komar, O.B. Serov, . "*Works on the holographic cinematography in the USSR*", SPIE 1183 p. 170-182 (1989).
3. . J.J. Lunazzi, "*Holophotography, a complete stereo technique*", Annual Meeting of the Society for Imaging Science and Technology-SPSE(IS&T), Boston-USA, 14-19.5.89, p.50.
4. J.J. Lunazzi, "*A new proposal for holography in white light*", "Holography '89" Intl. Meeting, Varna, Bulgaria, 22-25.5.89, p.8.
5. J.J. Lunazzi, personal communication to E. N. Leith, M. Baumstein, 1990.
6. J.J. Lunazzi, *"New possibilities in the utilisation of holographic screens"*, SPIE 1667(1992) p.289-293.
7. J.J. Lunazzi, P.M. Boone, "*One step technique for enlarging straddling holographic images by white-light projection onto a diffractive screen*", Optics Letters 19, 22, Nov.1994, p.1897-1899.
8. J.J. Lunazzi et al, "*Enlarging holograms under white light: 35mm format*", Proc. of the VI Symp. on Laser and Applications of the São Paulo State, São Carlos- SP - Brazil, 16-19.10.94, p.244.
9. J.J. Lunazzi, "*Enlarging Holograms Under White Light",* Proc. of the 17th Gral. Meeting of the International Commission for Optics, Taejeon, Korea, 19-23.08.96, SPIE V 2778 p.469-470.

---

J.J. Lunazzi,
Physics Institute
Campinas State University
C.P.6165
Campinas-SP- Brazil 13083-100
<lunazzi@ifi.unicamp.br>

<http://www.geocities.com/doctorlunazzi/jjl_i.htm>